\documentclass[preprint,aps,12pt,showpacs,nofootinbib,tightenlines]{revtex4}
\usepackage{amsmath}
\usepackage{amssymb}
\usepackage{epsfig}
\usepackage{graphicx}
\begin{document}
\preprint{NJNU-TH-2006-03}

\newcommand{\beq}{\begin{eqnarray}}
\newcommand{\eeq}{\end{eqnarray}}
\newcommand{\non}{\nonumber\\ }

\newcommand{\widet}[1]{ \widetilde{#1} }
\newcommand{\mw}{ M_W }
\newcommand{\tm}{ \widetilde{m} }
\newcommand{\tab}[1]{Table \ref{#1}}
\newcommand{\fig}[1]{Fig.\ref{#1}}
\newcommand{\real}{{\rm Re}\,}
\newcommand{\im}{{\rm Im}\,}
\newcommand{\calo}{ {\cal O} }

\newcommand{\smallsm}{{\scriptscriptstyle SM}}
\newcommand{\smallnp}{{\scriptscriptstyle NP}}

\def \epjc{  Eur. Phys. J. C }
\def \jpg{  J. Phys. G }
\def \mpla{ Mod.Phys.Lett. A }
\def \npb{  Nucl. Phys. B }
\def \plb{  Phys. Lett. B }
\def \prd{  Phys. Rev. D }
\def \prl{  Phys. Rev. Lett.  }
\def \pr{   Phys. Rep. }
\def \rmp{  Rev. Mod. Phys. }
\def \zpc{  Z. Phys. C  }

\title{$B\to X_s\gamma$, $X_s l^+ l^-$ decays and constraints on the mass insertion parameters
in the MSSM }
\author{ Zhenjun Xiao} \email{xiaozhenjun@njnu.edu.cn}
\author{Fengying Li}
\author{Wenjuan Zou}
\affiliation{Department of Physics and Institute of Theoretical Physics,
Nanjing Normal University, Nanjing, Jiangsu 210097, P.R.China}
\date{\today}
\begin{abstract}
In this paper, we study the upper bounds on the mass insertion parameters
$(\delta^{q}_{AB})_{ij}$ in the minimal supersymmetric standard model (MSSM).
We found that the information from the measured branching ratio of
$B \to X_s l^+ l^-$ decay can help us to improve the upper bounds on the mass insertions
parameters $\left ( \delta^{u,d}_{AB}\right )_{3j,i3}$. Some regions allowed by
the data of $Br(B \to X_s \gamma) $ are excluded by the requirement of a SM-like
$C_{7\gamma}(m_b)$ imposed by the data of $Br(B \to X_s l^+ l^-)$.
\end{abstract}

\maketitle

\section{Introduction}

As is well-known, the radiative $B \to X_s \gamma$ decay and the
semileptonic $B \to X_s l^+ l^-$ decay play an important role in the
precision test  of the standard model (SM) and the search for the new
physics beyond the SM. Although no evidence of the new physics
have been found now in experiments, one can put strong constraints on the
parameter space of various new physics model from currently
available data.

The minimal supersymmetric standard model (MSSM) is the general and most economical
low energy supersymmetric extension of the SM.  In order to find the
possible signals or hints of new physics from the date, various scenarios of the
MSSM are proposed by imposing different constrains on it \cite{ad99}.
Here, we use the pMSSM (phenomenological MSSM) model \cite{ad99} in our studies.

When calculating the new physics contributions to the flavor changing neutral current
(FCNC) processes induced by the loop diagrams involving the new particles, one needs
some kinds of model-independent parametrization of the FCNC SUSY contributions.
The mass insertion approximation (MIA) \cite{mia} is the best one of such kind
of parametrization methods. In the MIA, one chooses a basis for
the fermion  and sfermion states where all the couplings of these particles
to neutral gaugino fields are flavor diagonal, and leaves all the sources of flavor
violation inside the off-diagonal terms $\left ( \Delta^q_{AB}\right )_{ij}$
of the sfermion mass matrix, ie.,
\beq
\left ( M^2_{\widet{q}}\right )_{ij} = \widetilde{m}^2
 \delta_{ij} + \left ( \Delta^q_{AB}\right )_{ij}
 \eeq
where $\tm$ is the averaged squark mass, $A,B=(L,R)$, $q=u,d$, and the generation index
$i,j=(1,2,3)$.

As long as the off-diagonal terms is much smaller than the averaged squark mass,
the sfermion propagators can be expanded as a series in terms of
$\left ( \delta^q_{AB}\right )_{ij} = \left ( \Delta^q_{AB}\right )_{ij}/\tm^2$.
Under the condition of $\left ( \Delta^q_{AB}\right ) \ll \tm^2 $, one can just take
the first term of this expansion and translate the relevant experimental measurements
into upper bounds on these $\delta$'s.

According the helicity of the fermion partner, the squark mixings
can be classified into left- or right-handed (L or R) pieces:
\beq
\left ( \delta^q_{LL}\right )_{ij}, \quad \left (
\delta^q_{RR}\right )_{ij}, \quad \left ( \delta^q_{LR}\right
)_{ij}, \quad \left ( \delta^q_{RL}\right )_{ij}\;.
\label{eq:dij}
\eeq
The $LL$ and $RR$ mixings represent the chirality conserving
transitions in the left- and right-handed squars, while the  $LR$
and $RL$ mixings refer to the chirality flipping transitions. Up
to now, many interesting works have been done to draw constraints
on the parameter $\delta$'s from the known data. The  strong
constraints on the mass insertion $LL$ and $RR$, for example, are
obtained from the measured $\Delta M_K, \Delta M_{B_d}$ and
$\epsilon_K$ \cite{ggms96}, while the mass insertions in the $LR$
and $RL$ are constrained by the data of $Br(b \to s \gamma)$
\cite{ggms96,khalila}, the ratio $\epsilon'/\epsilon$ and the
electric dipole moment (EDM) of the neutron, electron and mercury
atom\cite{abel05}.

Very recently, the possible SUSY corrections to the
branching ratios and the CP-violating asymmetries of $B \to K\pi, K\eta^\prime$
and $\phi K$ decays have been studied, for example,  in
Refs.\cite{khalila,abel05,npb710}.
Here, the size and phase of the MIA parameters $\left ( \delta^q_{AB}\right )_{3j,i3}$
with $i,j=(1,2)$ play an impotent role in explaining those observed ``puzzles" in B
experiments or not.

The measured values of $Br(B\to X_s \gamma)$ and $Br(B \to X_s l^+ l^-)$
\cite{hfag} agree perfectly with the SM predictions \cite{gm01,jhep04},
which leads to strong constraints on various new physics models \cite{ghm05,newp}.
From the well measured $B \to X_s \gamma$ decays, the magnitude-but not the sign- of
the Wilson coefficient $C_{7\gamma}(m_b)$ is strongly constrained.
The sign of $C_{7\gamma}(m_b)$ and its absolute value, however, can be
determined from the precision measurements of the $B \to X_s l^+ l^-$ decays.
The latest Belle and BaBar measurements of the inclusive $B \to X_s l^+ l^-$
branching ratios indicated that the sign of $C_{7\gamma}(m_b)$
should be the same as the $C_{7\gamma}^{SM}(m_b)$ \cite{ghm05}.
Therefore, when calculating the
upper bounds of the mass insertions $\left ( \delta^q_{AB}\right )_{3j,i3}$
in the MSSM, one should not only consider the constraint coming from the
branching ratio of $B\to X_s\gamma$, but also the new information for
the sign of  $C_{7\gamma}(m_b)$ from the new measurements of
$Br(B \to X_S l^+l^-)$.

This paper is organized as follows. In Sec.\ref{sec:2}, we show
the formulaes needed to calculate the branching ratio of  $B\to
X_{s}\gamma$ decay at next-to-leading (NLO) order.
The new physics contributions to the relevant Wilson coefficients
are also given in this section. Then in Sec.\ref{sec:3},
by comparing the theoretical predictions with the data of $B \to X_s \gamma$ and $B \to X_s l^+ l^-$
decay we update the upper bounds on those mass insertion parameters.
We considered the cases of one mass insertion and two mass insertion approximation.
Summery is given in the last section.

\section{$Br(B\to X_s\gamma)$ in the SM and MSSM} \label{sec:2}

\subsection{$Br(B\to X_s\gamma)$ in the SM}

In the SM, the effective Hamiltonian for $b \to s \gamma$ at scale
$\mu \sim {\cal O}(m_b)$ reads \cite{gam96}
\beq
{\cal H}_{eff}=-\frac{G_{F}}{\sqrt{2}}V_{tb}V_{ts}^{*}
\left[{\sum_{i=1}^6}C_{i}(\mu)
O_{i}(\mu)+C_{7\gamma}(\mu)O_{7\gamma}(\mu)+C_{8g}(\mu)O_{8g}(\mu)
             \right]+h.c. ,
\label{eq:heff} \eeq where $V_{tb}V_{ts}^{*}$ is the products of
elements of the Cabbibo-Kabayashi-Maskawa (CKM) quark mixing
matrix \cite{ckm}. The definitions and the explicit expressions of
the operators $O_{i}$ $(i=1\sim 6, 7\gamma, 8g)$ and the
corresponding Wilson coefficients $C_i$ can be found in
Ref.\cite{gam96}. In the SM, the Wilson coefficients appeared in
Eq.~(\ref{eq:heff}) are currently known at next-to-leading order
(NLO) and can be found easily in Ref.\cite{gam96}.

At the lower energy scale $\mu_b\simeq {\cal O}(m_b)$, the Wilson
coefficients at NLO level can be formally decomposed as follows
\beq
C_{i}(\mu_b)=C_{i}^{(0)}(\mu_b)+\frac{\alpha_{s}(\mu_b)}{4\pi}C_{i}^{(1)}(\mu_b)
, \eeq where $C_{i}^{(0)}$ and $C_{i}^{(1)}$ stand for the LO and
NLO order part, respectively. Finally, the branching ratio
Br$(B\to X_s \gamma)$, conventionally normalized to the
semileptonic branching ratio Br$^{\rm exp}(B\to X_c
e\nu)=(10.64\pm 0.23)\%$, is given by~\cite{bg98,xiao04}
\beq
\rm{ Br}(B\to X_s\gamma) &=& \rm{ Br}^{\rm{exp}}(B\to X_c e \nu )
\frac{|V_{ts}^{*} V_{tb}|^2}{|V_{cb}|^2} \cdot \frac{6
\alpha_{em}}{\pi g(z) k(z)}\left [ |\overline{D}|^2 + A + \Delta
\right ]\, ,  \label{eq:brsm}
\eeq
with
\beq
\bar{D}(\mu_b)&=& C_{7\gamma}^{(0)}(\mu_b)+\frac{\alpha_s (\mu_b)}{4\pi}
C_{7\gamma} ^{(1)}(\mu_b)\non &&+\frac{\alpha_s (\mu_b)}{4\pi} \left \{
\sum_{i=1}^{8}C_i^{(0)}(\mu_b) \left[r_i(z)+\gamma_{i7}^{(0)}
\log{\frac{m_b}{\mu_b}}\right]- \frac{16}{3} C_{7\gamma}^{(0)}(\mu_b)\right \} \non
&=& C_{7\gamma}(\mu_b) + V(\mu_b),\label{eq:dbar} \\
A(\mu_b) &=&\frac{\alpha_s(\mu_b)}{\pi} \sum_{i,j=1;i \le j}^8 \,
  {\rm Re} \left\{ C_i^{0,\,{\rm eff}}(\mu_b) \,
            \left[ C_j^{0,\,{\rm eff}}(\mu_b)\right]^*
 \, f_{ij} \right\} \, \label{eq:amub}, \\
\Delta(\mu_b) &=& \frac{\delta_\gamma^{\smallnp}}{m_b^2} \left |C_7^{0,eff}(\mu_b)\right |^2
+ \frac{\delta_c^{\smallnp}}{m_c^2} {\rm Re}\left \{ \left [ C_7^{0,eff}(\mu_b)\right ]^*
\left [ C_2^{0,eff}(\mu_b)-\frac{1}{6}C_1^{0,eff}(\mu_b) \right ] \right \}
\label{eq:delta}
\eeq
with
\beq
\delta_{\gamma}^{\smallnp} &=& \frac{\lambda_1}{2} - \frac{9}{2}\lambda_2, \quad
\delta_{c}^{\smallnp} = - \frac{\lambda_2}{9}, \label{eq:dcnp}
\eeq
where $z=(m_c^{pole}/m_b^{pole})^2$, $\lambda_1=0.5\, {\rm GeV}^2$, $\lambda_2 = (m_{B^*}^2 - m_B^2)/4
= 0.12$ GeV$^2$ and $ m_b/2 \leq \mu_b \leq 2 m_b $.
The explicit  expressions for the Wilson coefficients $C_i^{(0)}$, $C_i^{(1)}$,  the
anomalous dimension matrix $\gamma$, together with the functions
$g(z)$, $k(z)$, $r_i(z)$ and $f_{ij}(\delta)$, can be found in
Refs.~\cite{bg98,xiao04,bsgnlo}. The term $A(\mu_b)$ describes the correction from the
bremsstrahlung process $b \to s\gamma g$, while the term $\Delta$ includes the non-perturbative
$1/m_b$ \cite{bsgnpmb} and $1/m_c$ \cite{bsgnpmc} corrections.

\begin{table}[thb]
\begin{center}
\caption{ The input parameters entering the calculation of $Br(B
\to X_s \gamma)$, their central values and errors. All masses are
in unit of GeV. We use $G_F=1.1664\times 10^{-5} \rm{GeV}$.
}\label{tab:input} \vspace{0.2cm}
\begin{tabular} {llllll}  \hline
$M_W$ &$M_Z$ & $m_t$ & $m_b$&$\alpha^{-1}$ & $\alpha_S(M_Z)$  \\ \hline
$80.42$ \hspace{1.5cm}&$91.188$ \hspace{1.5cm}& $175\pm 5$ \hspace{1.5cm}&
$4.8$\hspace{1.5cm}
&$137.036$ \hspace{1.5cm}& $0.118$ \hspace{1.5cm} \\ \hline
$BR_{SL}$ &$m_c/m_b$         & $\mu_b$
& $\sin^2{\theta_W}$&$\lambda_1$ & $\lambda_2$  \\ \hline
$0.1064$  &$0.29\pm 0.02 $ & $4.8^{+4.8}_{-2.4}$
& $0.23$            &$-0.5$      & $0.12$  \\ \hline
$A$      &$\lambda$ & $\bar{\rho}$       & $\bar{\eta}$
&$\left |\frac{V_{tb}V_{ts}}{V_{cb}}\right|^2$ &  \\ \hline
$0.854$  &$0.2196 $ & $0.20\pm 0.09$     & $0.33\pm 0.06$
&$0.97\pm 0.01$      &  \\ \hline
\end{tabular}\end{center}
\end{table}

Using above formulae and the input parameters as given in Table \ref{tab:input},
we find the SM prediction for the branching ratio
$Br(B\to X_s \gamma)$,
\beq
\rm Br^{\rm NLO}(B\to X_s\gamma)=(3.53\pm 0.30)\times 10^{-4}
\label{eq:bsgsm}
\eeq
where the main theoretical errors come from the uncertainties of the input parameters and have been added in
quadrature. By using the same input parameters, it is easy to find the numerical values of
$C_{7\gamma}(m_b)$, $V(m_b)$, $A(m_b)$ and $\Delta (m_b)$ as defined in Eqs.(\ref{eq:dbar}-\ref{eq:delta})
\beq
C_{7\gamma}(m_b)&=& -0.3052, \quad V(m_b)=-0.0257 -0.0156 I, \non
A(m_b) &=& 0.0033, \quad \Delta(m_b) = - 0.0010
\label{eq:c7mbv}
\eeq
One can see that it is the Wilson coefficient $C_{7\gamma}(m_b)$ who determines the branching ratio of
$B \to X_s \gamma$ decay, and other three terms are indeed very small.

\subsection{$Br(B\to X_s\gamma)$ in the MSSM}

At the leading order, the one-loop diagrams involving internal line SUSY particles
provide new physics contributions to the Wilson coefficients $C_{7\gamma}$, $C_{8g}$,
as well as the Wilson coefficients
$\widet{C}_{7\gamma}$ and $\widet{C}_{8g}$ of the new operators
$\widet{O}_{7\gamma}$ and $\widet{O}_{8g}$,
which have the opposite chirality with $O_{7\gamma}$ and $O_{8g}$ appeared
in Eq.(\ref{eq:heff}).
In the SM, the contributions from chiral-flipped operators $\widet{O}_{7\gamma,8g}$
are very small since they are strongly suppressed by a ratio $\calo (m_s/m_b)$.
In the MSSM, however, the contributions from the operators
$\widet{O}_{7\gamma}$ and $\widet{O}_{8g}$ may be not small in the SUSY models
with non-universal A-terms \cite{ggms96,npb710,npb353}.

At the one-loop level, there are four kinds of SUSY contributions to
the $b\to s$ transition,
depending on the virtual particles running in the penguin diagrams:
\begin{itemize}
\item[]
(i) the charged Higgs boson $H^{\pm}$ and up quarks $u,c,t$;

\item[]
(ii) the charginos $\tilde{\chi}^{\pm}_{1,2}$ and the
up squarks $\tilde{u}, \tilde{c},\tilde{t}$;

\item[]
(iii) the neutralinos $\tilde{\chi}^{0}_{1,2,3,4}$ and the
down squarks $\tilde{d}, \tilde{s},\tilde{b}$;

\item[]
(iv) the gluinos $\tilde{g}$ and the down squarks $\tilde{d}, \tilde{s},\tilde{b}$.

\end{itemize}

In general, the Wilson coefficients at $\mu_W \sim \mw$ after the inclusion of various
contributions can be expressed as
\beq
C_i(\mu_W) = C_{i}^{SM} + C_{i}^{H} +C_{i}^{\chi^+} +C_{i}^{\chi^0} +C_{i}^{\tilde{g}},
 \label{eq:cimuw}
\eeq
where $C_i^{SM}, C_{i}^{H}, C_{i}^{\chi^+}, C_{i}^{\chi^0}$ and $C_{i}^{\tilde{g}}$ denote
the Wilson coefficients induced by the penguin diagrams with the exchanges
of the gauge boson $W^\pm$, the charged Higgs $H^\pm$, the chargino $\chi^{\pm}_{1,2}$,
the neutralino $\chi^{0}_{1,2,3,4}$ and the gluino $\tilde{g}$, respectively
\footnote{The parameter $\lambda_t=V_{tb}V_{ts}^*$
in Eq.(10) of Ref.\cite{npb710} has been absorbed into the definition of $C^\chi_{i}$
and $C_i^{\tilde{g}}$ here. }

In principle, the neutrolino exchange diagrams involve the same
mass insertions as the gluino ones, but they are strongly
suppressed compared with the latter by roughly a ratio of
$\alpha/\alpha_s \approx 0.06$.
The charged Higgs contribution are proportional to the Yukawa
couplings of light quarks and relevant only for  a very small charged Higgs mass and
very large $\tan\beta$. Therefore, We will concentrate on the chargino and
gluino contributions only.
The SUSY contributions to the Wilson coefficients $C_{7\gamma,8g}$ and $\widet{C}_{7\gamma,8g}$
induced by various Feynman diagrams at the $m_W$ scale have been calculated and collected,
for example, in Refs.~\cite{ggms96,npb710,npb353,npb433}.

First, we consider the chargino contributions. As emphasized in
Ref.~\cite{npb710}, the chargino contributions induced by
magnetic-penguin, and chromomagnetic-penguin diagrams depend on
the up sector mass insertions $(\delta_{LL}^u)_{32,31}$ and
$(\delta_{RL}^u)_{32,31}$, while the $LR$ and $RR$ contributions
are suppressed by $\lambda^2$ or $\lambda^3$, where $\lambda=0.22$
is the Cabibbo mixing.

As for the mass spectrum of the squarks, the authors of Ref.\cite{npb710}
considered two cases: (a)  all squarks have the same mass $\tilde{m}$;
and (b) the stop-right $\tilde{t}_R$ is lighter than other squarks.
But the numerical results show that the difference between these two cases
are rather small. We therefor consider the first case only.

At the high energy scale $\mu_W \approx \mw$,
the chargino part of the Wilson coefficients, $C_{7\gamma}^{\chi}(m_W)$ and
$C_{8g}^{\chi}(m_W)$, are of the form \cite{npb710}
\beq
C_{7\gamma}^{\chi}(m_W)&=&  \lambda_t^{-1}\;\left \{  \left [ (\delta_{LL}^u)_{32} +
\lambda (\delta_{LL}^u)_{31} \right ] R_{\gamma}^{LL} + \left
[(\delta_{RL}^u)_{32} + \lambda (\delta_{RL}^u)_{31} \right ]\;
Y_t\, R_{\gamma}^{RL} \right \} , \label{eq:c7chi}\\
C_{8g}^{\chi}(m_W)&=& \lambda_t^{-1}\;\left \{ \left [ (\delta_{LL}^u)_{32} + \lambda
(\delta_{LL}^u)_{31} \right ] R_{g}^{LL}
+ \left [(\delta_{RL}^u)_{32}
+ \lambda (\delta_{RL}^u)_{31} \right ]\; Y_t\, R_{g}^{RL} \right \},
\label{eq:c8chi}
\eeq
with
\beq
R_{\gamma, g}^{LL}&=&\sum_{i=1}^2 |V_{i1}|^2\, x_{Wi}\, P_{\gamma,g}^{LL}(x_i)
- Y_b \sum_{i=1}^2 V_{i1} U_{i2}\, x_{Wi}\,
\frac{m_{\chi_i}}{m_b} P_{\gamma,g}^{LR}(x_i), \non
R_{\gamma, g}^{RL}&=& -\sum_{i=1}^2 V_{i1}V_{i2}^{\star}\,
x_{Wi}\,P_{\gamma,g}^{LL}(x_i),
\label{eq:rterms}
\eeq
and
\beq
P_{\gamma}^{LL}(x)&=&\frac{x \left ( -2-9x + 18 x^2-7x^3 +3 x (x^2-3)
\log[x] \right )}{9 (1-x)^5}, \non
P_{\gamma}^{LR}(x)&=& \frac{x \left (13-20 x+7 x^2+ (6+4 x-4 x^2)
\log[x] \right )}{6 (1-x)^4}, \non
P_{g}^{LL}(x)&=&\frac{x \left ( -1 + 9x + 9x^2 - 17x^3 + 6x^2 (3 + x)
\log[x] \right )}{12 (1 - x)^5}, \non
P_{g}^{LR}(x)&=& \frac{x \left (-1 - 4 x + 5 x^2 - 2 x (2 + x)
\log[x] \right )}{2 (1 - x)^4}.
\label{eq:pgg}
\eeq
where $Y_t= \sqrt{2} m_t /(v \sin\beta)$ and $Y_b=\sqrt{2} m_b \sqrt{1+ \tan^2\beta}/v$
are the Yukawa coupling of the top and bottom quark,
$x_{W_i}=m_W^2/m_{\chi_i}^2$, $x_{i}=m_{\chi_i}^2/\widetilde{m}^2$,
$\bar x_i =\widetilde{m}^2/m_{\chi_i}^2$ with $i=(1,2)$.
Finally, $U$ and $V$ in Eq.~(\ref{eq:rterms}) are the matrices that diagonalize chargino
mass matrix, which is defined by
\beq
U^{*}M_{\widetilde{\chi}^{+}}V^{-1}= diag\left ( m_{\widetilde{\chi}_{1}^{+}},
m_{\widetilde{\chi}_{2}^{+}} \right),
\eeq
with
\beq
M_{\widetilde{\chi}^{+}}=\left( {\begin{array}{*{20}c}
M_{2}&\sqrt{2}m_W\sin\beta\\\sqrt{2}m_W\cos\beta&\mu\\
\end{array}}\right).
\eeq
Here $M_2$ is the weak gaugino mass, $\mu$ is the supersymmetric
Higgs mixing term, and $\tan\beta =v_2/v_1$ is the ratio of the vacuum
expectation value (VEV) of the two-Higgs doublet. From above functions one can see that
\begin{itemize}
\item
The second term in $R_{\gamma}^{LL}$ and
$R_{g}^{LL}$ are enhanced by the large ratio $m_{\chi_i}/m_b \sim 30 -100$
and therefore provide a large chargino contribution to both $C_{7\gamma}$ and $C_{8g}$.

\item
For a large $\tan\beta$, say around 30 to 50, the Yukawa coupling $Y_b$ is
also large and leads to a further enhancement  to the $LL$ terms in
$C_{7\gamma}$ and $C_{8g}$.

\end{itemize}
In order to illustrate clearly the impact of the chargino
contributions to $B\to X_s\gamma$ process, it is very useful to
present the explicit dependence of the Wilson coefficients
$C_{7\gamma}(\mw)$ and $C_{8g}(\mw)$ on the relevant mass
insertions. For gaugino mass $M_2=200$ GeV, averaged squark mass
$\widetilde{m}=500$ GeV, $\mu = 300$ GeV and $\tan \beta =20$, we
obtain numerically
\beq
 C_{7\gamma}^{\chi}(\mw) &=& 0.411(\delta^u_{LL})_{31} +1.869(\delta^u_{LL})_{32}
 + 0.002(\delta^u_{RL})_{31} +  0.011(\delta^u_{RL})_{32}, \label{eq:chi1} \\
C_{8g}^{\chi}(\mw) &=& 0.104(\delta^u_{LL})_{31} + 0.475 (\delta^u_{LL})_{32}
+  0.001(\delta^u_{RL})_{31} + 0.004(\delta^u_{RL})_{32}, \label{eq:chi2}\\
\widetilde{C}_{7\gamma}^{\chi}(\mw) &=& 1.112 (\delta^u_{LR})_{31} + 5.062(\delta^u_{LR})_{32}
+ 0.009(\delta^u_{RR})_{31} +  0.042 (\delta^u_{RR})_{32}, \label{eq:chi3} \\
\widetilde{C}_{8g}^{\chi}(\mw) &=& 0.507 (\delta^u_{LR})_{31} + 2.309(\delta^u_{LR})_{32}
+  0.006(\delta^u_{RR})_{31} + 0.026 (\delta^u_{RR})_{32} . \label{eq:chi4}
\eeq
It is evident that the MIA parameter $(\delta^u_{LL})_{32}$( $ (\delta^u_{LR})_{32}$)
dominates the chargino contribution to $C_{7\gamma,8g}$ ($\widetilde{C}_{7\gamma,8g}$).
However, one should be
careful with this contribution since it is also the main
contribution to the $b\to s \gamma$, and stringent constraints on
$(\delta^u_{LL})_{32}$ are usually obtained, specially with large
$\tan \beta$. Finally, as expected from Eq.~(\ref{eq:rterms}),
only $LL$ contributions to $C^{\chi}_{7\gamma}$ and
$C^{\chi}_{8g}$ have strong dependence on the value of $\tan \beta$.
With $\tan \beta=40$, for instance, these contributions are
enhanced by a factor 2, while the $LR$ part of
$C^{\chi}_{7\gamma}$ and $C^{\chi}_{8g}$ are changing from the previous
values by less than $2\%$.

Now let us turn to the gluino contributions in the $b\to s$
transition. Its contribution to $C_{7\gamma}$ and $C_{8g}$ at scale $\mu_W $
are given by \cite{npb710}
\beq
C_{7\gamma}^{\widetilde{g}}&=&\frac{8\alpha_{s}\pi}{9\sqrt{2}G_{F}\tilde{m}^2 }
\, \lambda_t^{-1}\, \left[(\delta_{LL}^{d})_{23}M_{3}(x_{\tilde{g}})+(\delta_{LR}^{d})_{23}\frac{m_{\widetilde{g}}}{
     m_{b}}M_{1}(x_{\tilde{g}})\right],
  \label{eq:gluino1}   \\
C_{8g}^{\widetilde{g}}&=&\frac{\alpha_{s}\pi}{\sqrt{2}G_{F} \tilde{m}^2 }
 \, \lambda_t^{-1}\, \left[(\delta_{LL}^{d})_{23}\left({1\over 3}M_{3}(x_{\tilde{g}})
 +3M_{4}(x_{\tilde{g}})\right)\right.\non
&& \left.
+(\delta_{LR}^{d})_{23}\frac{m_{\widetilde{g}}}{m_{b}}\left({
1\over 3}M_{1}(x_{\tilde{g}})+3M_{2}(x_{\tilde{g}})\right)\right],
                       \label{eq:gluino2}\\
\widetilde{C}_{7\gamma}^{\widetilde{g}}&=&
\frac{8\alpha_{s}\pi}{9\sqrt{2}G_{F}\tilde{m}^2 }
 \, \lambda_t^{-1}\,    \left[(\delta_{RR}^{d})_{23}M_{3}(x_{\tilde{g}})+(\delta_{RL}^{d})_{23}
     \frac{m_{\widetilde{g}}}{m_{b}}M_{1}(x_{\tilde{g}})\right],
     \label{eq:gluino3}\\
\widetilde{C}_{8g}^{\widetilde{g}}&=&\frac{\alpha_{s}\pi}{\sqrt{2}G_{F} \tilde{m}^2}
 \, \lambda_t^{-1}\, \left[(\delta_{RR}^{d})_{23}\left({1\over 3}M_{3}(x_{\tilde{g}})+3M_{4}(x_{\tilde{g}})\right)
\right.\non
&& \left.
 +(\delta_{RL}^{d})_{23}\frac{m_{\widetilde{g}}}{m_{b}}\left(
 { 1\over 3}M_{1}(x_{\tilde{g}})+3M_{2}(x_{\tilde{g}})\right)\right].
\label{eq:gluino4}
\eeq
where $x_{\tilde{g}}= m_{\widetilde{g}}^2/\tilde{m}^2$ and the functions
$M_i(x)$ can be found in Ref.~\cite{npb710}.

Here in order to understand the impact of the gluino contributions to the
branching ratio of $b\to s \gamma$, we also present the explicit
dependence of the Wilson coefficients $C_{7\gamma, 8g}$ on the
relevant mass insertions. For $\widetilde{m}=500$Gev, and
$x=1$, we obtain
\beq
C_{7\gamma}^{\widetilde{g}}(\mw)& = &
- 0.049 (\delta^d_{LL})_{23} - 17.168 (\delta^d_{LR})_{23}, \label{eq:g1}
\\
C_{8g}^{\widetilde{g}}(\mw)     &= &  - 0.130(\delta^d_{LL})_{23}
- 64.379 (\delta^d_{LR})_{23} , \label{eq:g2}
\\
\widetilde{C}_{7\gamma}^{\widetilde{g}}(\mw)&= &
-0.049 (\delta^d_{RR})_{23} - 17.168 (\delta^d_{RL})_{23}, \label{eq:g3}
\\
\widetilde{C}_{8g}^{\widetilde{g}}(\mw) &\simeq &
-0.130 (\delta^d_{RR})_{23}  -  64.379(\delta^d_{RL})_{23}. \label{eq:g4}
\eeq
It is easy to see that the MIA parameter $(\delta^d_{LR})_{23}$ ( $(\delta^d_{RL})_{23}$ )
dominates the gluino contribution to $C_{7\gamma, 8g}^{\widetilde{g}}$
($\widetilde{C}_{7\gamma,8g}^{\widetilde{g}}$).

After the inclusion of the chargino and gluino contributions, the
total Wilson coefficients $C_{7\gamma}(M_W)$ and $C_{8g}(M_W)$
can be written as
\beq
C_{7\gamma(8g)}(M_W)&=&C_{7\gamma(8g)}^{SM}(M_W)+  C_{7\gamma(8g)}^\chi(M_W)
+C_{7\gamma(8g)}^{\widetilde{g}}(M_W), \label{eq:c7gtot}\\
\widetilde{C}_{7\gamma(8g)}(M_W)&=& \widetilde{C}_{7\gamma(8g)}^{\chi}(M_W) +
\widetilde{C}_{7\gamma(8g)}^{\widetilde{g}}(M_W).
\label{eq:c7gtot2}
\eeq
Since the heavy SUSY particles have been integrated out at the scale
$\mu_W =\mw $, the QCD running of the the Wilson coefficients $C_i(\mw)$ down to the
lower energy scale $\mu_b = {\cal O}(m_b)$ after including the new physics
contributions is the same as in the SM.

\section{the constraints on the mass insertion parameters}\label{sec:3}

In this section, we will update the constraints on the mass insertion parameters
$(\delta_{AB}^{u,d})_{ij}$ by considering the data of the branching ratios
for the $B \to X_s \gamma$ decay and the semileptonic $B \to X_s l^+ l^-$
decay.

For $B \to X_s \gamma$ decay, the new world average as given by Heavy Flavor
Averaging Group (HFAG) \cite{hfag} is
\beq
Br(B\to X_{s} \gamma) = \left ( 3.39 ^{+0.30}_{-0.27}\right ) \times
10^{-4}. \label{eq:bsg-exp}
\eeq
where the error is generally treated as $1\sigma$ error. From this result the
following bounds (at $3\sigma$ level) are obtained
\beq
2.53\times 10^{-4} < \rm Br(B\to X_{s} \gamma) < 4.34 \times
10^{-4} \label{eq:bound}.
\eeq
Here the experimental error at $3\sigma$ level has been  added in quadrature
with the theoretical error in the SM as given in Eq.(\ref{eq:bsgsm}).
From the bounds in Eq.(\ref{eq:bound}), the magnitude-but not the sign- of
the Wilson coefficient $C_{7\gamma}(m_b)$ can be strongly constrained. In the SM,
the allowed ranges for a real $C_{7\gamma}(m_b)$ are found to be
\beq
-0.360 &\leq & C_{7\gamma}(m_b)\leq -0.248, \label{eq:c7-b1}\\
0.454 & \leq & C_{7\gamma}(m_b)\leq 0.564, \label{eq:c7-b2}.
\eeq
But the second range where $C_{7\gamma}(m_b)$ is positive in sign is strongly
disfavored by the measured branching ratios of $B \to X_s l^+ l^-$ decays
\cite{ghm05}.

By using $M_2=200$ GeV, $\widetilde{m}=500$ GeV, $\mu = 300$ GeV, $\tan \beta =20$ and $x_{\tilde{g}}=1$,
we find the numerical result for the Wilson coefficient $C_{7\gamma}(m_b)$ after the inclusion
of new physics contributions from chargino and gluino penguins,
\beq
C_{7\gamma}(m_b) &=& -0.3052 -0.045 (\delta^d_{LL})_{23} +0.284 (\delta^u_{LL})_{31}
+1.295 (\delta^u_{LL})_{32}\non
&& -17.41 (\delta^d_{LR})_{23} -0.002 (\delta^u_{RL})_{31}  +0.008 (\delta^u_{RL})_{32},
\label{eq:c7mb}
\eeq
where the first term is the $C_{7\gamma}^{SM}(m_b)$ at the NLO level, other terms represent
the new physics contributions considered here.  Using the same input parameters, the chiral flipped
part $\widetilde{C}_{7\gamma}(m_b)$ is found to be
\beq
\widetilde{C}_{7\gamma}(m_b) &=& -0.045 (\delta^d_{RR})_{23} + 0.007 (\delta^u_{RR})_{31}
+0.030 (\delta^u_{RR})_{32}\non
&& -17.41 (\delta^d_{RL})_{23} +0.791 (\delta^u_{LR})_{31}  + 3.601 (\delta^u_{LR})_{32}.
\label{eq:c7mb2}
\eeq
One can see from Eqs.(\ref{eq:c7mb}) and (\ref{eq:c7mb2}) that
\begin{itemize}

\item[]
(i) for the MIA parameters $(\delta^u_{RL})_{31,32}$, $(\delta^d_{LL})_{23}$,
$(\delta^u_{RR})_{31,32}$ and $(\delta^d_{RR})_{23}$, there is no real constraint
\footnote{The term ``real constraint" means that the constraint of
$\left |\left ( \delta^{u,d}_{AB}\right )_{ij} \right | \ll 1 $. }
can be derived from the data of $B \to X_s \gamma$
and $B \to X_s l^+ l^-$ because of the tiny coefficients of these terms. We set these parameters zero
in the following analysis.

\item[]
(ii) the MIA parameters $(\delta^d_{LR})_{23}$ and $(\delta^u_{LL})_{31,32}$
($(\delta^u_{LR})_{31,32}$ and $(\delta^u_{RL})_{23}$ ) dominate the new physics
contributions to $C_{7\gamma}(m_b)$ ($\widetilde{C}_{7\gamma}(m_b)$).

\end{itemize}

When we take the new physics contributions into account,
the branching ratio of $B \to X_s \gamma$ decay can be written as
\beq
\rm{ Br}(B\to X_s\gamma) &=& \rm{ Br}^{\rm{exp}}(B\to X_c e \nu )
\frac{|V_{ts}^{*} V_{tb}|^2}{|V_{cb}|^2} \cdot \frac{6
\alpha_{em}}{\pi g(z) k(z)}\left [ \left |C_{7\gamma}(\mu_b) + V(\mu_b)\right |^2\right. \non
&& \left. + A(\mu_b)
+ \Delta(\mu_b) + \left | \widetilde{C}_{7\gamma}(\mu_b) \right |^2 \right ]\, ,
\label{eq:brnp2}
\eeq
where $C_{7\gamma}(\mu_b)$ includes the SM and the SUSY contributions as shown in eq.(\ref{eq:c7mb}),
while the last term refers to the new physics
contribution coming from the chiral-flipped operators $\widetilde{O}_{7\gamma}$ and $\widetilde{O}_{8g}$.
For the functions $V(\mu_b)$, $A(\mu_b)$ and $\Delta(\mu_b)$ in Eq.~(\ref{eq:brnp2}), the possible new physics
contributions to  these small high order quantities are also small and will be neglected in the following
numerical calculations.

Since the gluino and the chargino contributions are given in terms of the
MIA parameters of the up and down squark sectors, they are, in principle,
independent and could have constructive or destructive interference between
themselves or with the SM contribution. If we consider all MIA parameters in
Eqs.(\ref{eq:c7mb}) and (\ref{eq:c7mb2}) simultaneously,
no meaningful constraint can be obtained indeed. So we consider the cases of one
mass insertion and two mass insertion only:
\begin{itemize}
\item[]
(i) one mass insertion: the SM contribution and one non-zero MIA parameter are taken into account;

\item[]
(ii) two mass insertion: the SM contribution and two non-zero MIA parameters are taken into account;
\end{itemize}

\subsection{One mass insertion: gluino contributions}

We now consider the case of only one mass insertion parameter is non-zero at a time.
The MIA parameters are in general may be complex
and can be written as
\beq
\left ( \delta^q_{LR})_{ij} \right )= \left | \delta^q_{LR})_{ij}\right| e^{i\theta}
\eeq

We firstly consider the case of  $ (\delta^d_{LR})_{23}\neq 0$.
The MIA parameter $ (\delta^d_{LR})_{23}$ dominate the gluino
contribution to the Wilson coefficient $C_{7\gamma}(m_b)$, and therefore strong constraints on its
magnitude, as listed in Table \ref{tab:bound1}, can be derived by considering the data of
$Br(B \to X_s \gamma)$. In numerical calculations, we assume $\widetilde{m}=500$ GeV,
$x_{\tilde{g}}=0.3, 1, 3$,
and $\rm{arg}(\delta^d_{LR})_{23}=0, \pi/2$ and $\pi$, respectively.
In Table \ref{tab:bound1}, we also show the constraints on the magnitude of the parameter
$ (\delta^d_{RL})_{23}$.

One can see from Table \ref{tab:bound1} that the upper bounds on $\vert (\delta^d_{LR})_{23}\vert$ are
sensitive to the phase of this mass insertion. But the phase of the parameter
$ (\delta^d_{RL})_{23}$ is still free parameter.
Since the parameter $(\delta^d_{RL})_{23}$ appears in $\widetilde{C}_{7\gamma}(m_b)$ only,
which does not mix with $C_{7\gamma}(m_b)$.

\begin{table}[tb]
\caption{Upper bounds on MIA parameters $\vert (\delta^d_{LR})_{23}\vert$ and
$\vert ( \delta^d_{RL} )_{23}\vert$ from the data of $Br(B\to X_s \gamma)$
for $\widetilde{m}=500$ GeV, $x_{\tilde{g}} = 0.3,~1,~3$
and $\rm{arg}(\delta^d_{LR})_{23}=0$ (a), $\pi/2$ (b) and $\pi$ (c),
respectively.}
\label{tab:bound1}
\begin{center}
\begin{tabular}{c|c|c|c}
\hline \hline x & 0.3 & 1 & 3\\
\hline
& (a) 0.0024 & (a) 0.0021 & (a)0.0014\\
$ \left | (\delta^d_{LR})_{23} \right|$&
(b) 0.0099 & (b)0.0083&(b)0.0054\\
    & (c) 0.0033 & (c) 0.0030 & (c) 0.0019 \\ \hline
$\left | (\delta^d_{RL})_{23}\right|$ &0.0102 &0.0107&0.0103\\
\hline\hline
\end{tabular}
\end{center}
\end{table}

\subsection{One mass insertion: Chargino contributions}

Now we consider the chargino contribution as the dominant SUSY
effect to $b\to s\gamma$. From Eqs.~(\ref{eq:c7mb}) and Eq.~(\ref{eq:c7mb2}),
one can see easily that the $LL$ and $LR$ sector mass insertions give the dominant contribution
to the Wilson coefficients $C_{7\gamma}(m_b)$ and $\widetilde{C}_{7\gamma}(m_b)$.

From the data of $Br(B \to X_s \gamma)$ as given in Eq.~(\ref{eq:bound}) and the requirement of a
SM-like $C_{7\gamma}(m_b)$ as indicated by the measured branching ratios of $B \to X_s l^+ l^-$ decay,
the upper bounds on $(\delta^u_{LL})_{31,32}$ and
$(\delta^u_{LR})_{31,32}$ as the function of the gaugino mass $M_2$ are given in Table.~\ref{tab:bound2}.
In the numerical calculation, we use $\widetilde{m}=500$ GeV, $\mu=\pm 300$ GeV, and $\tan\beta=20$.
For larger values of $\tan\beta$, the upper bounds as listed in Table \ref{tab:bound2} will be scaled by a
factor of $20/\tan\beta$.

\begin{table}[tb]
\caption{Upper bounds on $\vert (\delta^u_{LL})_{31,32}\vert$ and
$\vert (\delta^u_{LR})_{31,32}\vert$ from the data of $Br(B\to X_s \gamma)$ and
$Br(B\to X_s l^+l^-)$ for $\widetilde{m}=500$ GeV, $\mu=300$ GeV,
$\tan\beta=20$ and the phase $\theta=0$ (a), $\pi/2$ (b) and $\pi$ (c), respectively.}
\label{tab:bound2}
\begin{center}
\begin{tabular}{c|c|c|c|c|c|c}\hline\hline
& \multicolumn{3}{|c|}{$\mu=300$}& \multicolumn{3}{|c}{$\mu=-300$}\\ \hline
$M_2$& 200 &400 &600   &200&400&600\\
\hline
&(a)0.185&(a)0.267&(a)0.365 &(a)0.135&(a)0.191&(a)0.264        \\
$\left | (\delta^u_{LL})_{31}\right|$
&(b)0.618&(b)0.894 &(b)1.230&(b)0.532&(b)0.765 &(b)1.05\\
&(c)0.127 &(c)0.185 &(c)0.255 &(c)0.192 &(c)0.282 &(c)0.382\\  \hline
&(a)0.042 &(a)0.057 &(a)0.082 &(a)0.028 &(a)0.042 &(a)0.058 \\
$\left | (\delta^u_{LL})_{32}\right|$
&(b)0.138 &(b)0.195&(b)0.270 &(b)0.116 &(b)0.168&(b)0.230\\
&(c)0.028 & (c)0.042&(c)0.056 &(c)0.042 & (c)0.061&(c)0.084\\ \hline
$\left | (\delta^u_{LR})_{31}\right|$
& 0.201 & 0.197& 0.195 & 0.204 & 0.199& 0.197\\ \hline
$\left | (\delta^u_{LR})_{32}\right|$
& 0.044 & 0.044& 0.043 & 0.045 & 0.044& 0.043\\ \hline
\hline
\end{tabular}
\end{center}
\end{table}

\begin{figure}[tbp]
\begin{center}
\vspace{-1cm}
\includegraphics[width=12cm]{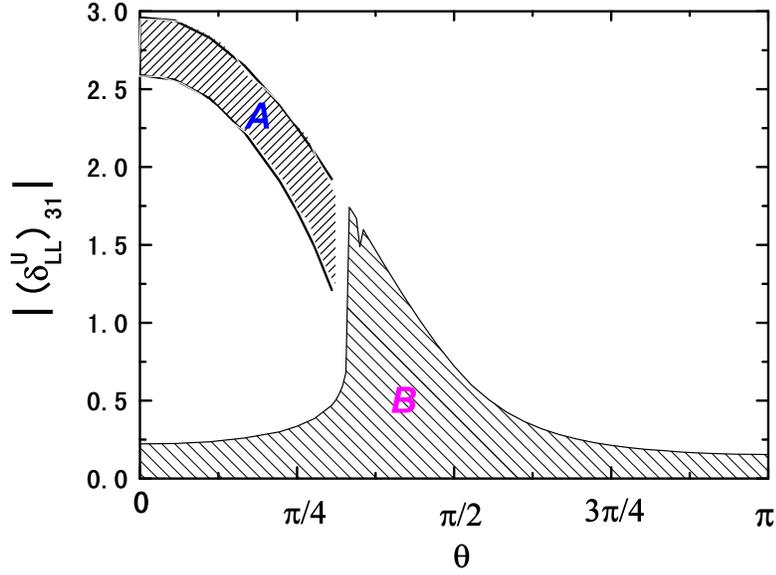}
\caption{Plot of $\vert (\delta^u_{LL})_{31}\vert$ as a function
of its phase $\theta$. Here both A and B regions are allowed by the data of  $Br(B \to X_s \gamma)$,
but the region A is excluded by the requirement of a SM-like $C_{7\gamma}(m_b)$.}
\label{fig:fig1}
\vspace{-1cm}
\end{center}
\end{figure}

\begin{figure}[tbp]
\begin{center}
\includegraphics[width=12cm]{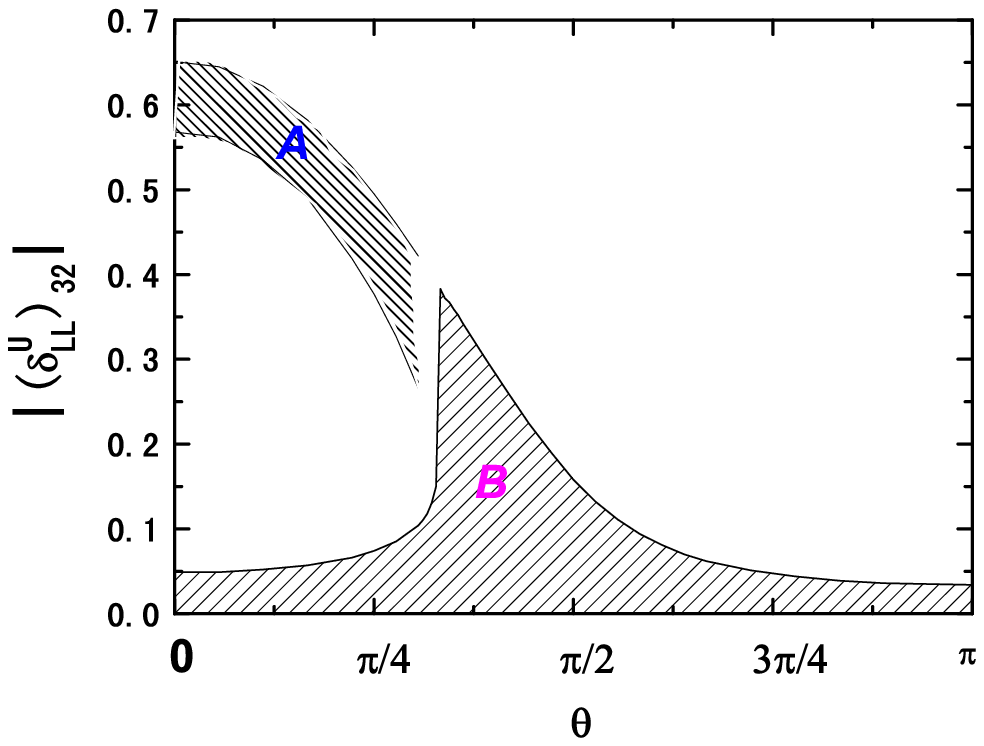}
\caption{Plot of $\vert (\delta^u_{LL})_{32}\vert$ as a function
of its phase $\theta$. Here both A and B regions are allowed by the data of  $Br(B \to X_s \gamma)$,
but the region A is excluded by the requirement of a SM-like $C_{7\gamma}(m_b)$.}
\label{fig:fig2}
\vspace{-1cm}
\end{center}
\end{figure}

By taking into account the data of $Br(B \to X_s l^+ l^-)$, the constraints on the
MIA parameters become more stronger than those obtained by considering the data of $Br(B \to X_s \gamma)$
only. In Fig.~\ref{fig:fig1} and \ref{fig:fig2}, we show the the phase dependence of the bounds on
$ \vert (\delta^u_{LL})_{31}\vert$ and $ \vert (\delta^u_{LL})_{32}\vert$ by assuming $M_2=200$ GeV,
$\widetilde{m}=500$ GeV, $\mu=300$ GeV and $\tan\beta=20$.
Here both the shadow regions of A and B are allowed by the data of $Br(B \to X_s \gamma)$,
but the region A is excluded by considering the additional restrictions coming from the data
of  $Br(B \to X_s l^+ l^-)$.

\begin{table}[tb]
\caption{Upper bounds on $\vert (\delta^d_{LL})_{13,23}\vert$ obtained from
Table \ref{tab:bound2} and Eq.~(\ref{eq:d31}) and (\ref{eq:d32}). }
\label{tab:bound3}
\begin{center}
\begin{tabular}{c|c|c|c|c|c|c}\hline\hline
& \multicolumn{3}{|c|}{$\mu=300$}& \multicolumn{3}{|c}{$\mu=-300$}\\ \hline
$M_2$& 200 &400 &600   &200&400&600\\\hline
&(a)0.176&(a)0.255&(a)0.347 &(a)0.129&(a)0.182&(a)0.251        \\
$\left | (\delta^d_{LL})_{13}\right|$
&(b)0.588&(b)0.851 &(b)1.17 &(b)0.507&(b)0.728 &(b)0.999\\
&(c)0.121 &(c)0.176 &(c)0.243 &(c)0.183 &(c)0.269 &(c)0.364 \\  \hline
&(a)0.083&(a)0.116&(a)0.162 &(a)0.058&(a)0.084&(a)0.116        \\
$\left | (\delta^d_{LL})_{23}\right|$
&(b)0.274&(b)0.391 &(b)0.540&(b)0.233&(b)0.336 &(b)0.461\\
&(c)0.056 &(c)0.083 &(c)0.112 &(c)0.084 &(c)0.123 &(c)0.168\\  \hline
\hline
\end{tabular}
\end{center}
\end{table}

Furthermore, because of the $SU(2)$ gauge invariance the soft
scalar mass $M_{Q}^2$ is common for the up and down sectors.
Therefore, one can get the following relations between the up and
down type mass insertions
 \beq
(\delta^d_{LL})_{ij} &=& \left[V_{CKM}^+~ (\delta^u_{LL})~
V_{CKM} \right]_{ij}~.
\eeq
For the elements $ij=31,32$, we have
\beq
(\delta^d_{LL})_{31} &=&(\delta^u_{LL})_{31} - \lambda (\delta^u_{LL})_{32}+
\mathcal{O}(\lambda^2)~, \label{eq:d31}\\
(\delta^d_{LL})_{32} &=&(\delta^u_{LL})_{32} +\lambda (\delta^u_{LL})_{31}+
\mathcal{O}(\lambda^2)~. \label{eq:d32}
\eeq
Consequently, the upper bounds on $\vert(\delta^u_{LL})_{31,32}\vert$
can be conveyed to a constraint on $\vert(\delta^d_{LL})_{31,32}\vert$ which
equals to $\vert(\delta^d_{LL})_{13,23}\vert$, due to the hermiticity
of $(M_{D}^2)_{LL}$. From the bounds as given in Table \ref{tab:bound2} and the relation in
eqs.~(\ref{eq:d31}) and (\ref{eq:d32}), the upper bounds on
$\vert(\delta^d_{LL})_{13,23}\vert$ can be obtained, as listed in Table \ref{tab:bound3},
where the terms proportional to $\lambda^2$ or higher powers are neglected.
These are the strongest constraints one may
obtain on $\vert(\delta^d_{LL})_{13,23}\vert$, and therefore it
should be taken into account in analyzing the ``LL" part of the
gluino contribution to the $b \to s$ and $b \to d$ transitions.

\subsection{Two mass insertion}

Finally we consider the scenario in which two mass insertion parameters are non-zero
at a time. Because of the interference between different terms, the situation become more
complicated than the cases of one mass insertions.
By assuming $M_2=200$ GeV, $\widetilde{m}=500$ GeV, $x_{\tilde{g}}=1$ and $\tan\beta=20$, we
consider only two typical cases:
\begin{itemize}
\item[]
Case-A: $(\delta^u_{LL})_{32}$ and $ (\delta^u_{LR})_{23}$ in $C_{7\gamma}(m_b)$ are non-zero, i.e.
\beq
C_{7\gamma}(m_b)&=& -0.3052 -17.41 (\delta^d_{LR})_{23} + 1.295 (\delta^u_{LL})_{32}, \non
\widetilde{C}_{7\gamma}(m_b)&=& 0. \label{eq:c7mb21}
\eeq

\item[]
Case-B:  $(\delta^d_{RL})_{23}$ and $ \delta^u_{LR})_{32}$ in $\widetilde{C}_{7\gamma}(m_b)$
are non-zero, i.e.
\beq
C_{7\gamma}(m_b)&=& -0.3052, \non
\widetilde{C}_{7\gamma}(m_b)&=& -17.41 (\delta^d_{RL})_{23} + 3.601 (\delta^u_{LR})_{32}.
\label{eq:c7mb22}
\eeq
\end{itemize}

\begin{figure}[tb]
\begin{center}
\vspace{-1cm}
\includegraphics[width=11cm]{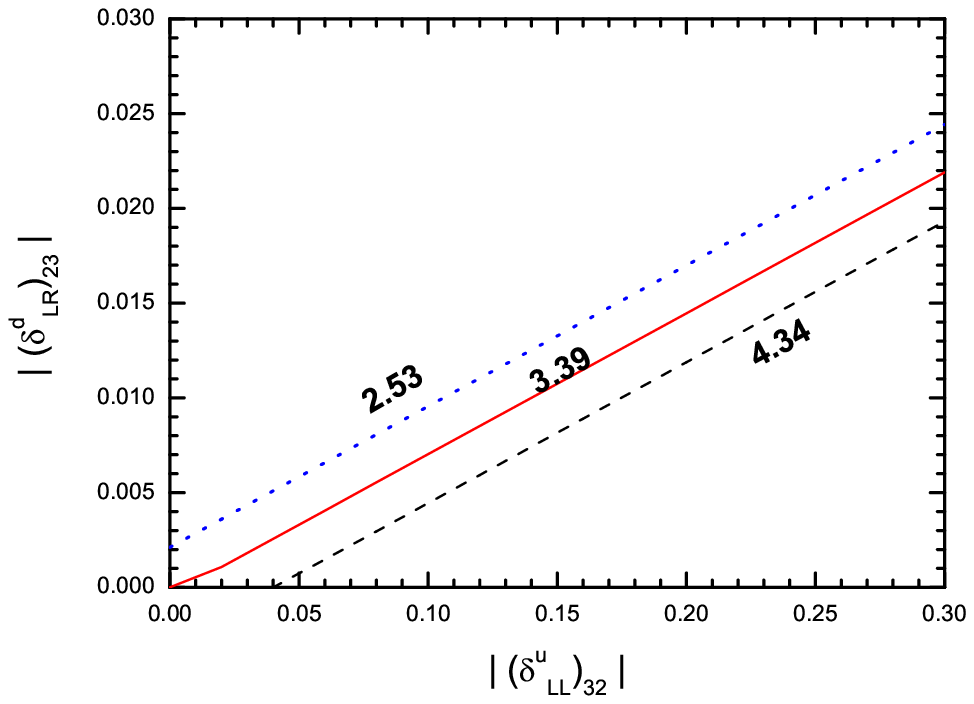}
\caption{Contour plot of $Br(B \to X_s \gamma) \times 10^{4} $ as a function of
$|(\delta^u_{LL})_{32}|$ and $|(\delta^d_{LR})_{23}|$ with a same zero phase.}
\label{fig:fig3}
\vspace{-1cm}
\end{center}
\end{figure}

\begin{figure}[tb]
\begin{center}
\includegraphics[width=11cm]{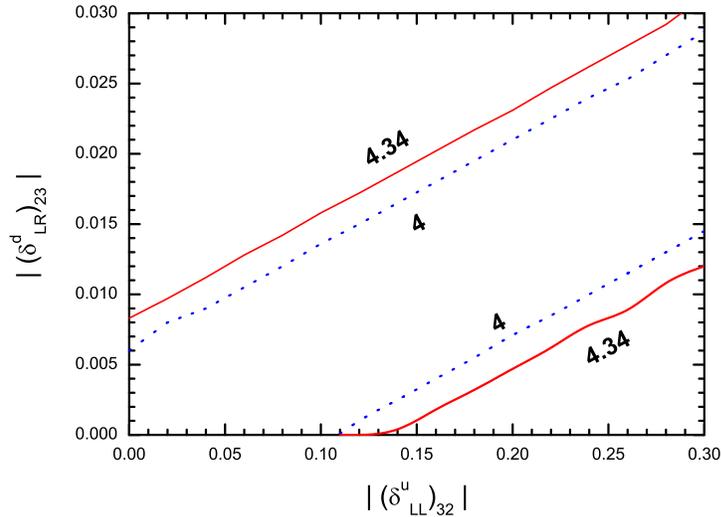}
\caption{Contour plot of $Br(B \to X_s \gamma)  \times 10^{4} $ as a function of
$|(\delta^u_{LL})_{32}|$ and $|(\delta^d_{LR})_{23}|$ with a same phase of $\theta=\pi/2$.}
\label{fig:fig4}
\vspace{-1cm}
\end{center}
\end{figure}

\begin{figure}[tb]
\begin{center}
\vspace{-1cm}
\includegraphics[width=11cm]{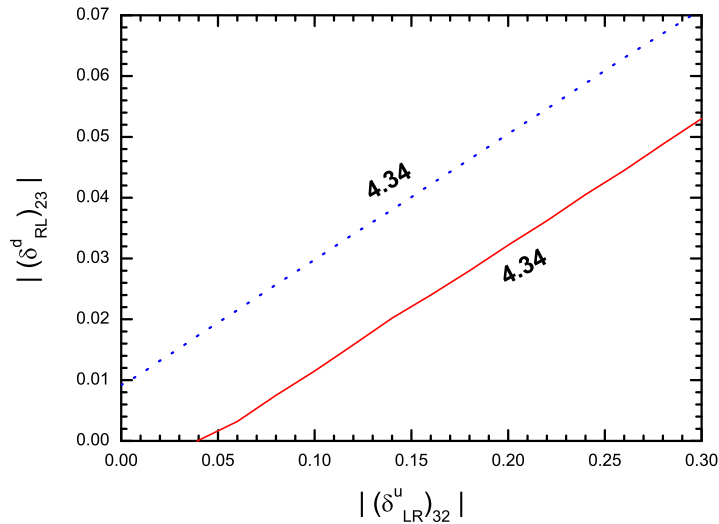}
\caption{Contour plot of $Br(B \to X_s \gamma)  \times 10^{4} $ as a function of
$|(\delta^u_{LR})_{32}|$ and $|(\delta^d_{RL})_{23}|$ with the same zero phase .}
\label{fig:fig5}
\vspace{-1cm}
\end{center}
\end{figure}

We firstly consider the case A, where the SM and the chargino contributions are taken into account.
In Fig.~\ref{fig:fig3}, we show the contour plot of  $Br(B \to X_s \gamma)$ (in unit of $10^{-4}$)
as a function of $|(\delta^u_{LL})_{32}|$ and $|(\delta^d_{LR})_{23}|$ with a same zero phase.
Where the dotted, dashed line shows the lower and upper bounds of the branching ratios as given
in Eq.~(\ref{eq:bound}): $ 2.53\times 10^{-4} \leq Br(B \to X_s \gamma) \leq 4.34\times 10^{-4}$,
while the solid line refers to the central value of the data.
The region between the dotted and dashed lines
are allowed by the data of both $B \to X_s \gamma$ and $B \to X_s l^+ l^-$ decay.
The Fig.~\ref{fig:fig4} shows the same contour plot as Fig.~(\ref{fig:fig3}), but for
a non-zero phase of $\theta=\pi/2$. The region between two solid lines
is allowed by the data of both $B \to X_s \gamma$ and $B \to X_s l^+ l^-$ decay.

We now consider the case B, where the SM and the gluino contributions are taken into account.
In Fig.~\ref{fig:fig5}, we show the contour plot of  $Br(B \to X_s \gamma)$ (in unit of $10^{-4}$)
as a function of $|(\delta^u_{LR})_{32}|$ and $|(\delta^d_{RL})_{23}|$ with a same zero phase.
The region between the dotted and solid lines are allowed by the data of both
$B \to X_s \gamma$ and $B \to X_s l^+ l^-$ decay.

\section{summary}

In this paper, by comparing the theoretical predictions with the corresponding measured
branching ratios of $B \to X_s \gamma$ and $B \to X_s l^+ l^-$ decays,
we make an update for the upper bounds on the mass insertions parameters
$(\delta^{u,d}_{AB})_{ij}$ appeared in the MSSM.

From the well measured $B \to X_s \gamma$ decays, the strong constraints on the MIA parameters
can be obtained. From the latest Belle and BaBar measurements of the inclusive
$B \to X_s l^+ l^-$ branching ratios, further constraints can be derived.
We found that the information from the measured branching ratio of
$B \to X_s l^+ l^-$ decay can help us to improve the upper bounds on the mass insertions
parameters $\left ( \delta^q_{AB}\right )_{3j,i3}$ in the MSSM.

We focus on the possible large new physics contributions to the Wilson coefficients $C_{7\gamma,8g}$
and $\widetilde{C}_{7\gamma,8g}$ coming from chargino and gluino penguin diagrams.
Throughout our analysis, the SM contributions are always taken into account.
From the numerical calculations, we found the following points
\begin{enumerate}
\item
In the one mass insertion approximation, strong upper bounds on the mass insertion parameters
$\left ( \delta^u_{LL}\right )_{31,32}$, $\left ( \delta^d_{LL}\right )_{13,23}$,
$\left ( \delta^d_{LR}\right )_{23}$, $\left ( \delta^u_{LR}\right )_{31,32}$ and
$\left ( \delta^d_{RL}\right )_{23}$  are obtained, as collected in Tables \ref{tab:bound1}-\ref{tab:bound3}.

\item
As shown explicitly in Figs.~\ref{fig:fig1} and \ref{fig:fig2}, the region A allowed by the data of
$Br(B \to X_s \gamma) $ is excluded by the requirement of a SM-like $C_{7\gamma}(m_b)$ imposed
by the data of $Br(B \to X_s l^+ l^-)$.

\item
Under two mass insertion approximation, strong upper bounds on $\left ( \delta^u_{LL}\right )_{23}$,
and $\left ( \delta^d_{LR}\right )_{23}$, and $\left ( \delta^u_{LR}\right )_{23}$ and
$\left ( \delta^d_{RL}\right )_{23}$ can also be obtained, as shown in Figs.~\ref{fig:fig3}-\ref{fig:fig5}.

\end{enumerate}

\begin{acknowledgments}

This work is partly supported  by the National
Natural Science Foundation of China under Grant No.10275035, 10575052,
and by the Research Foundation of Nanjing Normal University under Grant No.~214080A916.

\end{acknowledgments}


\end{document}